\documentclass[graybox]{svmult}
\usepackage{mathptmx}       
\usepackage{helvet}         
\usepackage{courier}        
\usepackage{makeidx}         
\usepackage{graphicx}        
\usepackage{multicol}        
\usepackage{amsmath,amssymb,amsfonts}        
\makeindex             
\usepackage{subcaption}

\newcommand{\sfrac}[2]{{\textstyle\frac{#1}{#2}}}

\title*{Self-similar rupture of a thin viscous filament in a Hele-Shaw cell}
\author{Rodney Dharma \and Michael C. Dallaston
\and Marco A. Fontelos \and \\ Bronwyn H. Bradshaw-Hajek
 \and Scott W. McCue \thanks{
This report presents the results of a project undertaken by the first author at the MATRIX Workshop \textit{Slow Viscous Flows with Interfaces}, 7--18 July 2025, under the supervision of the other authors.
}}
\institute{R. Dharma \at School of Mathematics \& Statistics, University of Melbourne, Parkville VIC 3010 \at \email{r.dharma1@student.unimelb.edu.au}
\and M.C. Dallaston, S.W. McCue \at School of Mathematical Sciences, Queensland University of Technology, Brisbane QLD 4001 \at \email{michael.dallaston@qut.edu.au}, \email{scott.mccue@qut.edu.au}
\and M.A. Fontelos \at Instituto de Ciencias Matem\'{a}ticas, Campus de Cantoblanco, 28006 Madrid, Spain \at \email{marco.fontelos@icmat.es}
\and B.H. Bradshaw-Hajek \at UniSA STEM, The University of South Australia, Mawson Lakes SA 5095 \at \email{bronwyn.hajek@unisa.edu.au}
}
\authorrunning{R. Dharma \and M.C. Dallaston \and M.A. Fontelos \and B.H. Bradshaw-Hajek \and S.W. McCue}

\begin{document}

\maketitle

\section{Introduction}

There are many mathematical and experimental studies of the motion of a viscous fluid sandwiched between two parallel plates.  This set-up is often referred to as a Hele-Shaw cell \cite{heleshaw1898}.  Possibly the most interesting class of Hele-Shaw flows occurs when there is a sharp interface between the viscous fluid and the surrounding air (or surrounding less viscous fluid), which may or may not be stable, depending on whether the air is displacing the fluid or vice verca~\cite{Morrow2021,Paterson1981,saffman1958}.  In this report, we focus on the situation in which there is a thin filament or thread of viscous fluid in a Hele-Shaw cell that is forced to become even thinner at a neck region and subsequently pinches off or ruptures.

\begin{figure}
\centering
\includegraphics[width=0.7\textwidth]{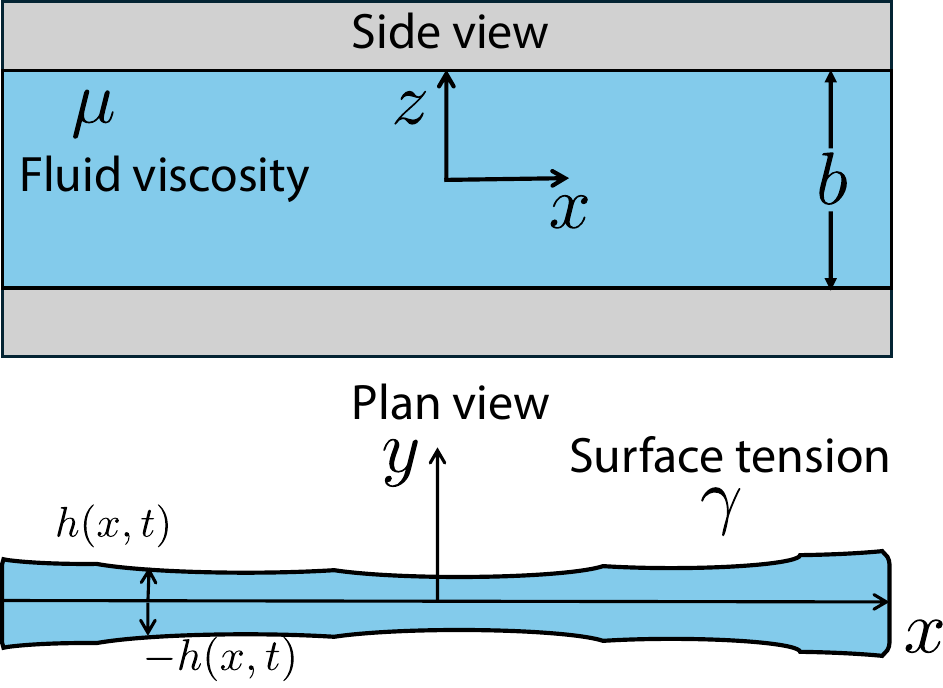}
\caption{A schematic of the filament in a Hele-Shaw cell.  (a) A plan view of the domain showing the thin filament that is much longer in the $x$-direction than its thickness in the $y$-direction.  (b) A side view of the Hele-Shaw cell with two flat parallel plates separated by a distance $b$.}
\label{fig:schematic}
\end{figure}

A very simple lubrication-type model for the evolution of a thin filament in a Hele-Shaw cell is
\begin{equation}
h_t+\frac{b^2\gamma}{12\mu}(hh_{xxx})_x=0,
\label{eq:pdedimensional}
\end{equation}
where $h(x,t)$ is (half of) the width of the filament, $\mu$ is fluid viscosity, $\gamma$ is the surface tension and $b$ is the gap between the parallel plates in the Hele-Shaw apparatus.  We are using subscripts in this equation to denote partial derivatives.  A schematic of this scenario is included in figure~\ref{fig:schematic}, showing both a top view in (a) and a side view in (b).  In short, the derivation of the lubrication model (\ref{eq:pdedimensional}) starts by assuming the $x$-component of velocity, $u$, can be approximated by averaging the Stokes flow equations over the small gap between the plates to give a Darcy-type law $u=-(b^2/12\mu)p_x$, where the pressure in the filament is related to the interface curvature $\kappa\approx h_{xx}$ via $p=-\gamma\kappa$.  Together, to leading order in the lubrication approximation, we combine $u=(b^2/12\mu)\gamma h_{xxx}$ with conservation of mass $h_t+(uh)_x=0$ to give (\ref{eq:pdedimensional}) (see \cite{Bertozzi1994,constantin1993,dupont1993}, for example).  We note in passing that this model requires the shape of the filament to be symmetric about the centreline.

We shall discuss various aspects of the (dimensionless version of the) model (\ref{eq:pdedimensional}) in some detail, but three obvious initial observations are worth noting.  First, (\ref{eq:pdedimensional}) is a (nonlinear fourth-order) diffusion equation, which reflects the tendency of surface tension to flatten out any nonuniformities in filament thickness.  Second, the diffusion is degenerate in the sense that the mobility $h$ vanishes in the limit $h\rightarrow 0^+$.  Thus, if the filament is forced to develop a neck (with a local minimum at $x=x_{\mathrm{min}}(t)$) that is decreasing in size, the dynamics near the neck are driven by the product of $h$ and its fourth derivative, since
$$
h_t \bigg|_{x=x_{\mathrm{min}}(t)} = - \frac{b^2\gamma}{12\mu}(hh_{xxxx})\bigg|_{x=x_{\mathrm{min}}(t)}.
$$
This means that for the filament to rupture ($h_{\mathrm{min}}\rightarrow 0^+$ as $t\rightarrow t_0^-$), there must be a singularity in the fourth-derivative, $h_{xxxx}$, in the limit.  Third, the partial differential equation (pde) (\ref{eq:pdedimensional}) admits a four-parameter Lie transformation group with the associated linearly independent operators
\begin{equation}
\Gamma_1=\frac{\partial}{\partial x}, \quad
\Gamma_2=\frac{\partial}{\partial t}, \quad
\Gamma_3=h\frac{\partial}{\partial h}-t\frac{\partial}{\partial t}, \quad
\Gamma_4=p\,h\frac{\partial}{\partial h}+x\frac{\partial}{\partial x}+(4-p)\, t\frac{\partial}{\partial t},
\label{eq:operators}
\end{equation}
where $p$ is a constant.  The fourth of these is equivalent to the statement that (\ref{eq:pdedimensional}) is invariant under the stretching transformation
\begin{equation}
x=\lambda \hat{x}, \quad t=\lambda^{4-p}\hat{t}, \quad h=\lambda^p \hat{h}.
\label{eq:stretching}
\end{equation}
As a consequence, the exists the possibility of similarity solutions of (\ref{eq:pdedimensional}) of the form $h=(t_0-t)^{p/(4-p)}f((x-x_0)/(t_0-t)^{1/(4-p)})$ for some $t_0>0$ and free parameter $p$, as we shall explore.

With these observations in mind, if we start out with a slightly perturbed interface, so that the thread is almost flat in shape, we would expect the diffusive nature of the model (\ref{eq:pdedimensional}) to kill off all modes of perturbation in such a way that the interface levels off and becomes flatter over time.  Indeed, this turns out to be true if we impose periodic boundary conditions.  On the other hand, if the two ends of the filament are pinned in place (at different fixed values of $x$) and fluid is withdrawn from each end with a sufficiently high outward flux, then the loss of fluid causes the filament to thin and ultimately to rupture at a finite time.  This type of filament rupture due to the model (\ref{eq:pdedimensional}) has been studied at length, most notably in the 1990s by Constantin et al.~\cite{constantin1993}, Dupont et al.~\cite{dupont1993} and Bertozzi and co-workers \cite{almgren1996,bertozzi1996,Bertozzi1994}.

In this report, we shall revisit the rupture problem for (\ref{eq:pdedimensional}), which, surprisingly, is still generating new and interesting research questions \cite{king2025}.  We continue in section~\ref{sec:numerical}, where we illustrate three typical solution behaviours of (\ref{eq:pdedimensional}) for appropriate combinations of initial and boundary conditions that include $h(-L,t)=h(L,t)=h_0$ for constants $h_0>0$, $L>0$.  These are infinite-time levelling solutions and those that appear to approach symmetric and asymmetric rupture in a finite time.  For the symmetric rupture case, we demonstrate in section~\ref{sec:similarity} how rupture is described by a similarity solution of the second kind (consistent with (\ref{eq:operators})--(\ref{eq:stretching})), which itself is governed by a nonlinear boundary value problem that can be solved numerically.  Surprisingly, there does not seem to be examples in the literature of time-dependent numerical solutions of (\ref{eq:pdedimensional}) evolving towards a second-kind similarity solution in the rupture limit.
In section~\ref{sec:tfe}, we briefly frame our results in a broader perspective by considering the thin film equation \cite{bertozzi1996,Bertozzi1994,myers1998,Smyth1988}
\begin{equation}
h_t+(h^mh_{xxx})_x = 0,
\label{eq:tfe}
\end{equation}
which for $m=1$ is a dimensionless version of (\ref{eq:pdedimensional}).
We present a type of bifurcation diagram that characterises symmetric similarity solutions of (\ref{eq:tfe}) as a function of $m$.  These results confirm those published very recently by King \& Bowen~\cite{king2025}.
Finally, we close in section~\ref{sec:discussion} with a discussion of our work and future research directions.

\section{Numerical solutions}\label{sec:numerical}

For definiteness, we shall consider the dimensionless version of (\ref{eq:pdedimensional}), namely
\begin{equation}
h_t+(hh_{xxx})_x=0,
\label{eq:nondimensional}
\end{equation}
on the interval $-1<x<1$, subject to the boundary conditions
\begin{equation}
h(-1,t)=h(1,t)=1, \quad t>0,
\label{eq:bcs1}
\end{equation}
\begin{equation}
h_{xx}(-1,t)=h_{xx}(1,t)=P(t), \quad t>0.
\label{eq:bcs2}
\end{equation}
Here, (\ref{eq:bcs1}) fixes the thickness of the filament at the end points of the domain, while (\ref{eq:bcs2}) provides a driving pressure field that acts to remove fluid from the filament.  For the case $P(t)\equiv 0$, together (\ref{eq:bcs1})--(\ref{eq:bcs2}) are equivalent to imposing periodic boundary conditions with no fluid loss. Otherwise, we shall employ a version of $P(t)$ taken from Dupont et al.~\cite{dupont1993}, as outlined below.

For our numerical investigation, it is worth noting that if $P(t)\rightarrow P_\infty$ as $t\rightarrow\infty$, then for $P_\infty<2$, the problem (\ref{eq:nondimensional})--(\ref{eq:bcs2}) has a steady-state solution $h=1-\sfrac{1}{2}P_\infty (1-x^2)$.  Thus, to drive the system to rupture, one option is to impose a pressure $P(t)$ such that $P_\infty>2$.

\subsection{Numerical scheme}

We simulate solutions of (\ref{eq:nondimensional})--(\ref{eq:bcs2}) using a method-of-lines approach. The spatial domain $\Omega = (-1, 1)$ is discretised into $N + 2$ grid points, where $N > 5000$ is chosen to ensure good accuracy. Second-order finite-difference approximations are applied to convert the pde (\ref{eq:nondimensional}) into a system of nonlinear ordinary differential equations (odes), which are then solved using MATLAB's ODE15s implicit numerical stiff solver.

For ease of notation, the grid points are defined as:
\[
x_i \triangleq \frac{2i}{N + 1} - 1, \quad i=1,2,\ldots,N,
\]
with $x_0 = -1$ and $x_{N+1} = 1$ representing the endpoints of the domain $\Omega$. The film height at each grid point is denoted as $h_i(t) = h(x_i, t)$. Using finite-difference approximations for the first and third spatial derivatives, the resulting system of odes can be expressed as:
$$
\frac{\mathrm{d}h_i}{\mathrm{d}t} = \frac{q_{i-1} - q_{i+1}}{2 \Delta x_i},
\quad
    q_i = h_i \bigg(\frac{h_{i+2} - 2h_{i+1} + 2h_{i-1} - h_{i-2}}{2 (\Delta x_i)^3}\bigg),
$$
where $q_i$ represents the numerical flux, and $\Delta x_i$ is the grid spacing.
To impose the boundary conditions (\ref{eq:bcs1}), we set $h_0 = h_{N+1} = 1$. Special care is required to impose the boundary pressure conditions (\ref{eq:bcs2}), as these correspond to constraints on the second derivatives
$\partial^2 h_0/\partial x^2$ and $\partial^2 h_{N+1}/\partial x^2$, which are not explicitly included in the discretised equations. These conditions are implemented by modifying the numerical flux equations for $i \in \{1, N_x\}$ as follows:
\begin{align*}
    q_1 &= h_1 \bigg(\frac{h_1 + h_3 - 2h_2}{2 \Delta x^3} - \frac{P(t)}{2 \Delta x}\bigg), \\[6pt]
    q_N &= -h_N \bigg(\frac{h_N + h_{N-2} - 2h_{N-1}}{2 \Delta x^3} - \frac{P(t)}{2 \Delta x}\bigg).
\end{align*}
Here, $P(t)$ represents the pressure term at the boundaries in (\ref{eq:bcs2}). These modifications ensure that the boundary pressure conditions are correctly incorporated into the numerical scheme.

\subsection{Exponential decay}

The first, and most straightforward, example is to treat (\ref{eq:nondimensional})--(\ref{eq:bcs2}) with $P(t)\equiv 0$, together with an initial condition of the form $h(x,0)=1+\epsilon\cos(n\pi x)$, where $\epsilon$ is a constant whose magnitude is less than unity.  Under the assumption $|\epsilon|\ll 1$, we can linearise (\ref{eq:nondimensional}) to show that $h\sim 1+\epsilon\,\mathrm{e}^{\lambda_n t}\cos(n\pi x)$, where $\lambda_n=-n^4\pi^4$.  In other words, under periodic boundary conditions, the evolution of a slightly perturbed filament interface effectively proceeds as if the nonlinear model (\ref{eq:nondimensional}) is its linear cousin $h_t+h_{xxxx}=0$.

\begin{figure}
\centering
\includegraphics[width=0.9\textwidth]{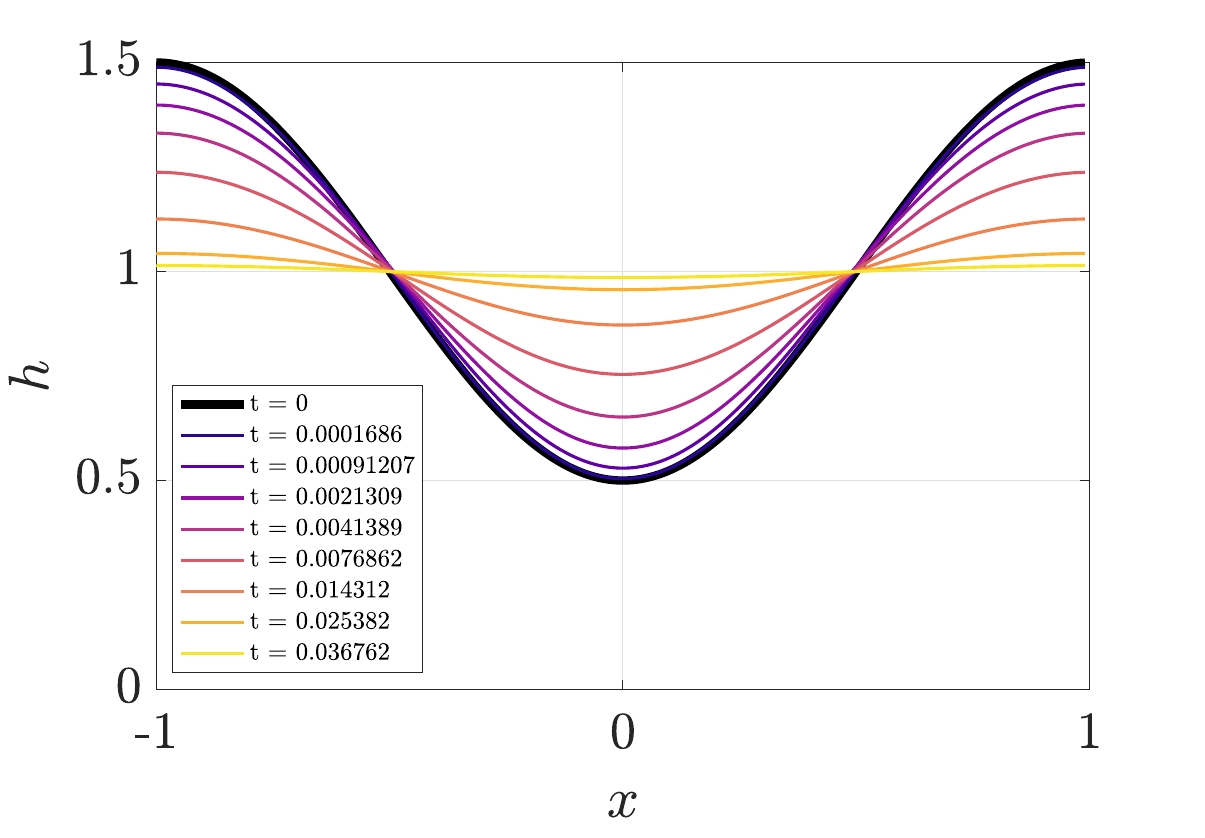}
\caption{Evolution of the filament interface for the initial condition $h(x,0)=1-\frac{1}{2}\cos(\pi x)$ with $P(t)\equiv 0$.  The initial condition is indicated by a thick curve, while the profiles for positive times are presented as thin curves.}
\label{fig:decay}
\end{figure}

To demonstrate one case, we set $h(x,0)=1-\frac{1}{2}\cos(\pi x)$ and solve (\ref{eq:nondimensional})--(\ref{eq:bcs2}) numerically with $P(t)\equiv 0$.  The results are presented in figure~\ref{fig:decay}.  Clearly the numerical solution is tracking the prediction from linear stability theory very closely, as expected.

\subsection{Symmetric rupture at a single point}

\begin{figure}
\centering
\includegraphics[width=0.9\textwidth]{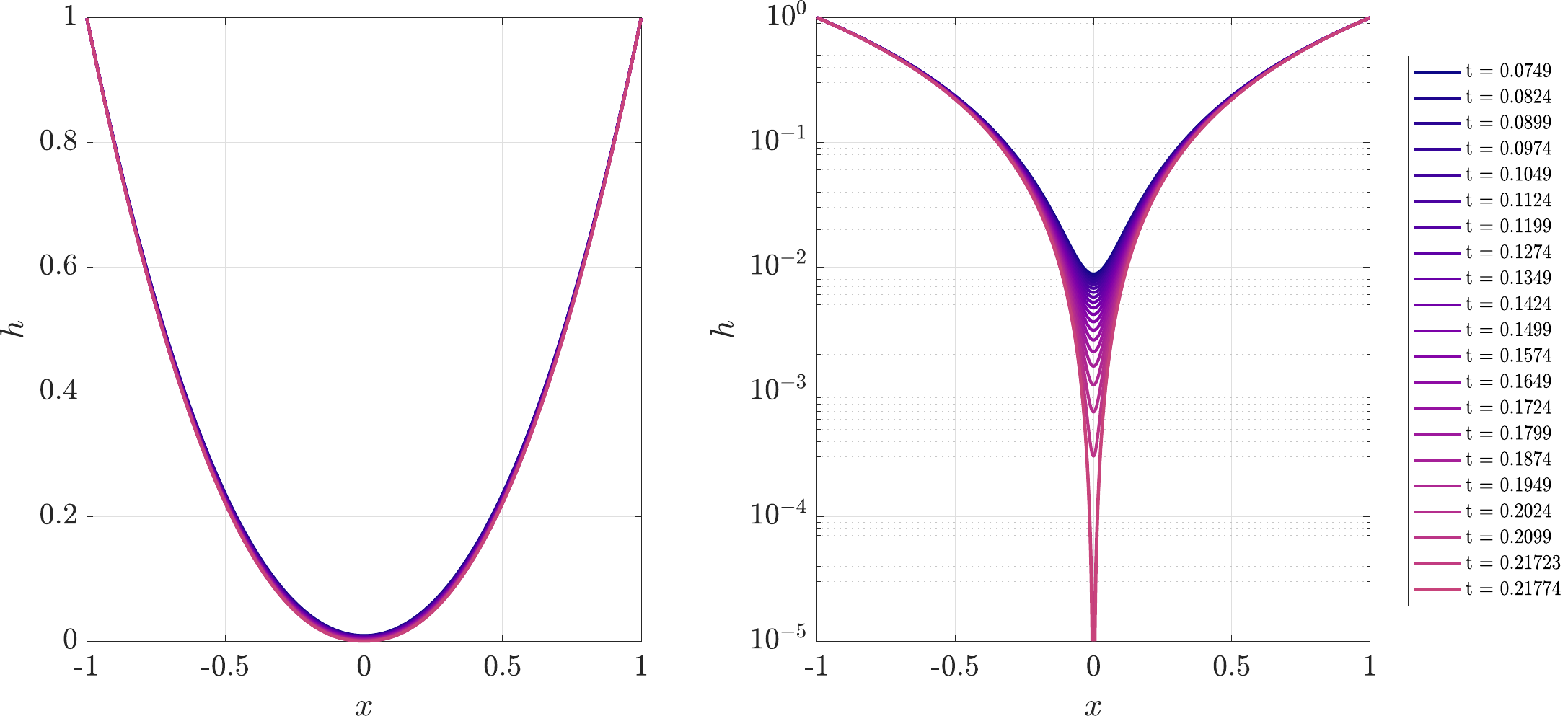}
\caption{Evolution of the filament interface for the initial condition (\ref{eq:IC_dupont}) with imposed pressure (\ref{eq:pressure}), where $p_{0} = 127/64$, $p_{f} = 2.5$ and $t_s=0.05$.  Both (a) and (b) show the same solutions; the only difference is the profiles $h(x,t)$ in (b) are plotted on a logarithmic scale.}
\label{fig:symmetricpde1}
\end{figure}

\begin{figure}
\centering
\includegraphics[width=0.9\textwidth]{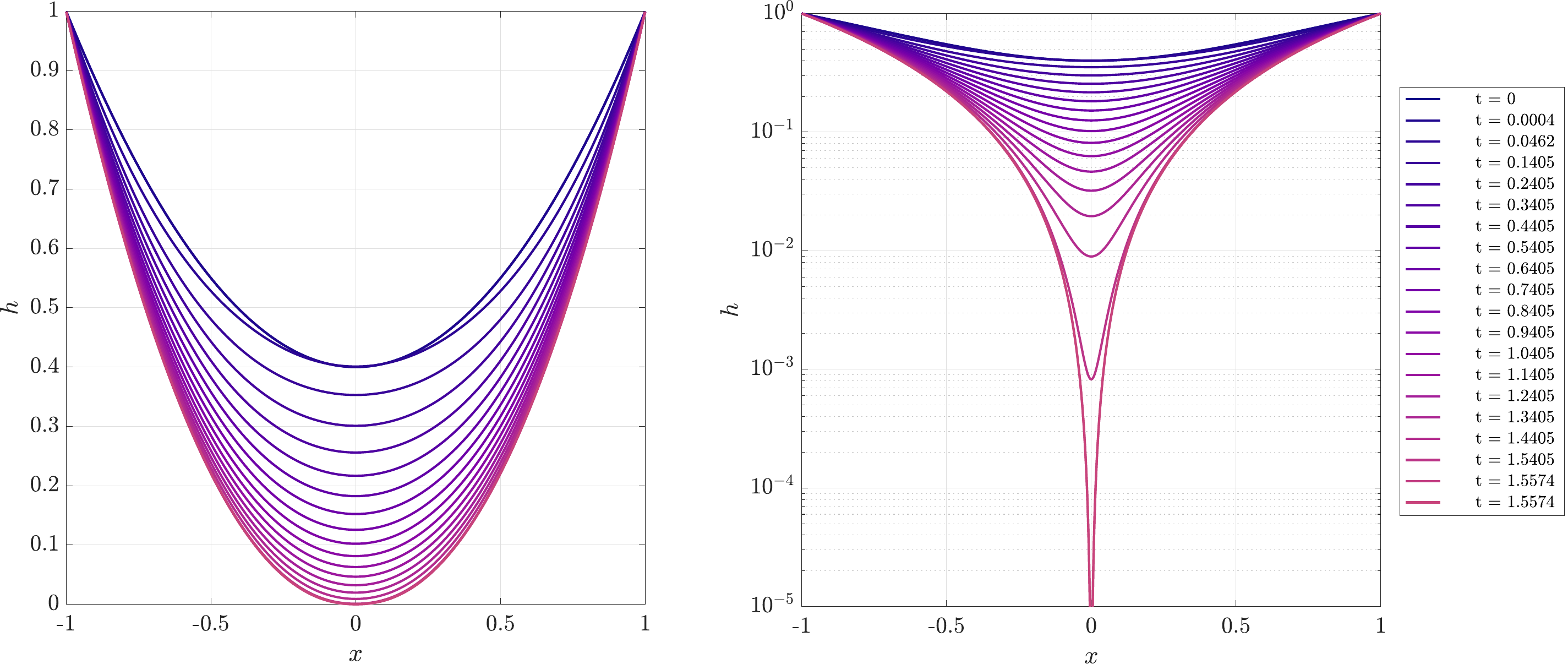}
\caption{Evolution of the filament interface for the initial condition (\ref{eq:IC_dupont}) with imposed pressure (\ref{eq:pressure}), where $p_{0} = 5/4$, $p_{f} = 2.5$ and $t_s=0.05$.}
\label{fig:symmetricpde2}
\end{figure}

Following in the spirit of Dupont et al.~\cite{dupont1993}, we consider (\ref{eq:nondimensional})--(\ref{eq:bcs2}) with the initial condition
\begin{equation}
h(x,0) = 1-\frac{1}{2}p_0(1-x^2),
\label{eq:IC_dupont}
\end{equation}
and impose the boundary pressure
\begin{equation}
P(t) =
\begin{cases}
p_{0} + (p_{f} - p_{0}) \left(3 - \dfrac{2t}{t_s}\right) \left(\dfrac{t}{t_s}\right)^{2}, & t < t_s, \\[1em]
p_{f}, & t \ge t_s.
\end{cases}
\label{eq:pressure}
\end{equation}
This pressure function has the properties that it starts at $P(0)=p_0$ and transitions to $P(t_s)=p_f$ at some intermediate time, where it remains thereafter.
We plot in figure~\ref{fig:symmetricpde1} a solution for $p_{0} = 127/64$, $p_{f} = 2.5$ and $t_s=0.05$.  In this case, the viscous filament begins with a significant neck that is already much thinner than the filament at the end points.  As such, given the driving force $P(t)$ that acts to remove fluid from the filament, the interface is drawn down and ultimately ruptures at a finite time $t=t_0$.  We include a similar plot in figure~\ref{fig:symmetricpde2}, except this time we change $p_0$ to be $p_{0} = 5/4$.  In this case, initially the filament is not particularly thin, but again goes on to rupture in finite time in a similar way to figure~\ref{fig:symmetricpde1}.

To clarify the near-rupture behaviour, we have shown interface profiles in figures~\ref{fig:symmetricpde1} and \ref{fig:symmetricpde2} using both primitive variables in (a) and using a log-scale for $h$ in (b).  Given the scales involved, it is difficult to appreciate the evolution in (a), especially in figure~\ref{fig:symmetricpde1}.  On the other hand, in (b) we see the filament thinning and approaching a rupture point at some finite time.

\begin{figure}
\centering
\includegraphics[width=0.8\textwidth]{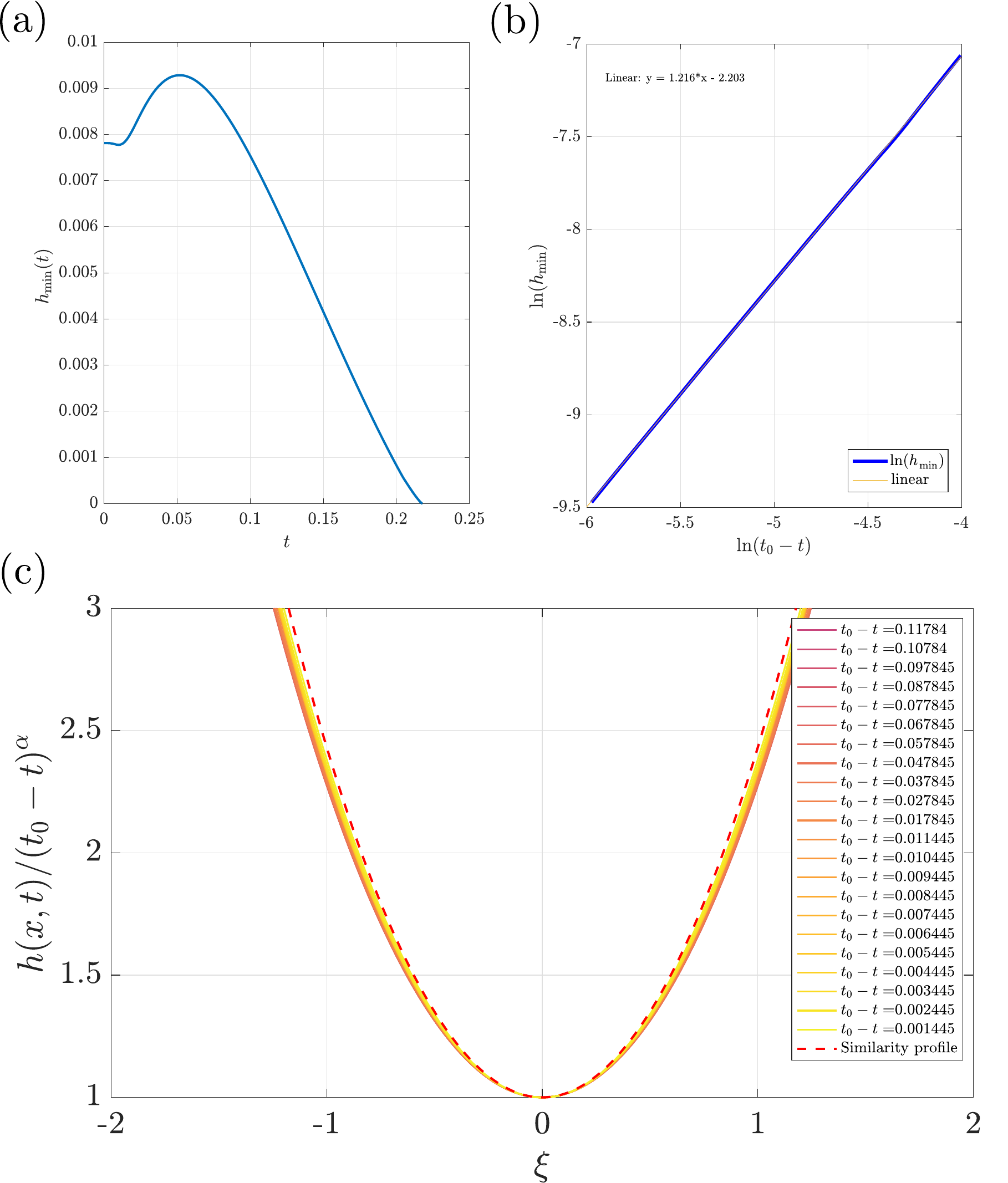}
\caption{Further results from the simulation shown in figure~\ref{fig:symmetricpde1}.  (a) A plot of the minimum thickness $h_{\mathrm{min}}(t)$ versus time, which is used to predict the rupture time $t_0$.  (b) The same data plotted on a log-log plot, which is used to predict the similarity parameter $\alpha\approx 1.22$.  (c)  Given the numerically estimated $t_0$ and the predicted value $\alpha=1.236$ from (\ref{eq:odef})--(\ref{eq:farfieldf}), rescaled profiles of the filament interface, $h(x,t)/(t_0-t)^\alpha$, are plotted against the similarity variable $\xi=x/(t-t_0)^{(1+\alpha)/4}$ using thin curves.  Also included as a red dashed curve is the relevant symmetric similarity solution $f(\xi)$ which is computed directly from the boundary value problem (\ref{eq:odef})--(\ref{eq:farfieldf}).}
\label{fig:symmetricode}
\end{figure}

\begin{figure}
\centering
\includegraphics[width=0.9\textwidth]{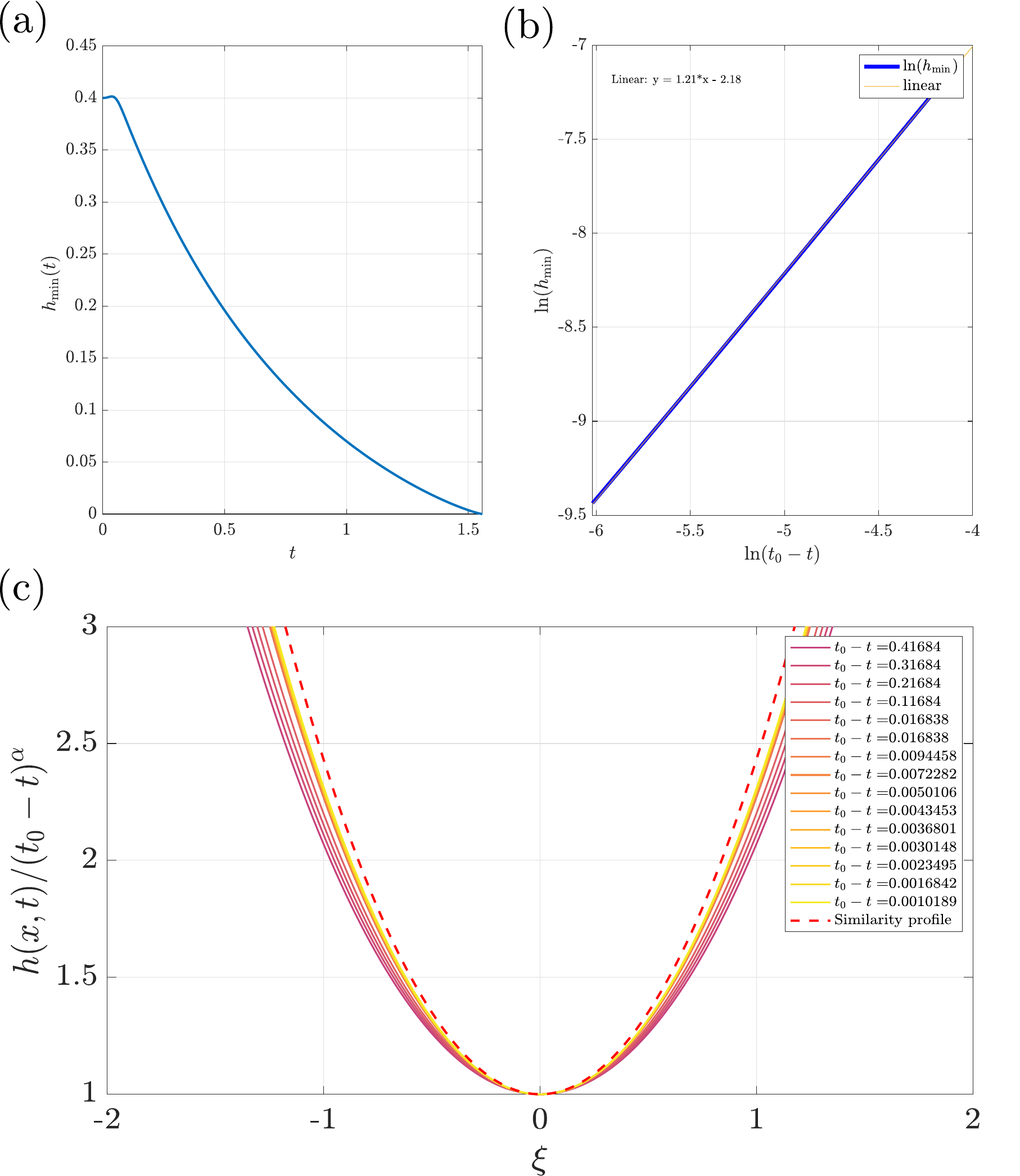}
\caption{Further results from the simulation shown in figure~\ref{fig:symmetricpde2}.  (a) A plot of the minimum thickness $h_{\mathrm{min}}(t)$ versus time, which is used to predict the rupture time $t_0$.  (b) The same data plotted on a log-log plot, which is used to predict the similarity parameter $\alpha\approx 1.21$.  (c)  Given the numerically estimated $t_0$ and the predicted value $\alpha=1.236$ from (\ref{eq:odef})--(\ref{eq:farfieldf}), rescaled profiles of the filament interface, $h(x,t)/(t_0-t)^\alpha$, are plotted against the similarity variable $\xi=x/(t-t_0)^{(1+\alpha)/4}$ using thin curves.  Also included as a red dashed curve is the relevant symmetric similarity solution $f(\xi)$ which is computed directly from the boundary value problem (\ref{eq:odef})--(\ref{eq:farfieldf}).}
\label{fig:symmetricode2}
\end{figure}

We are particularly interested in the behaviour near the rupture time.  One important measure is the minimum filament thickness $h_{\mathrm{min}}(t)$, which we can track numerically.  In figure~\ref{fig:symmetricode}(a), we plot $h_{\mathrm{min}}(t)$ versus time $t$ for the example shown in figure~\ref{fig:symmetricpde1}.  We see that initially the filament thickness increases, but then decreases until rupture.  We can extrapolate this data to estimate the rupture time $t_0$.  Using this value of $t_0$, we plot $\ln (h_{\mathrm{min}})$ versus $\ln(t_0-t)$ in figure~\ref{fig:symmetricode}(b).  Using interval $-6<\ln(t_0-t)<-4$, we can fit a line to the data to estimate the slope, which turns out to be roughly 1.22.  (This will be our numerical estimate for the parameter $\alpha$ defined below in (\ref{eq:similarityansatz})).  In a similar way, we have included a plot of $h_{\mathrm{min}}(t)$ versus $t$ in figure~\ref{fig:symmetricode2}(a) that corresponds to the solution shown in figure~\ref{fig:symmetricpde1}, together with a plot of $\ln (h_{\mathrm{min}})$ versus $\ln(t_0-t)$ in figure~\ref{fig:symmetricode2}(b).  This time the rough estimate for the slope is roughly 1.21.  We shall return to figures~\ref{fig:symmetricode} and \ref{fig:symmetricode2} in section~\ref{sec:similarity}.

\subsection{Asymmetric rupture at single point}

\begin{figure}
\centering
\includegraphics[width=0.9\textwidth]{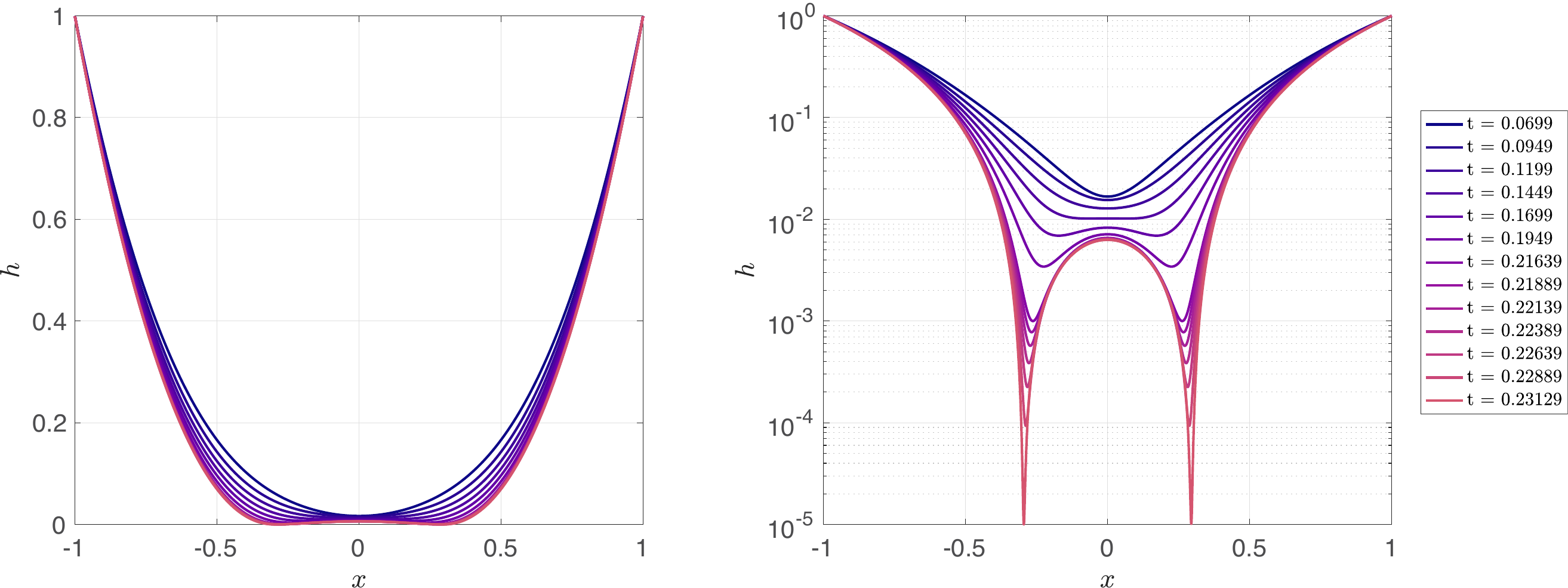}
\caption{Evolution of the filament interface for the initial condition (\ref{eq:IC_dupont}) with imposed pressure (\ref{eq:pressure}), where $p_{0} = 127/64$, $p_{f} = 5$ and $t_s=0.05$.}
\label{fig:asymmetricpde1}
\end{figure}

\begin{figure}
\centering
\includegraphics[width=0.9\textwidth]{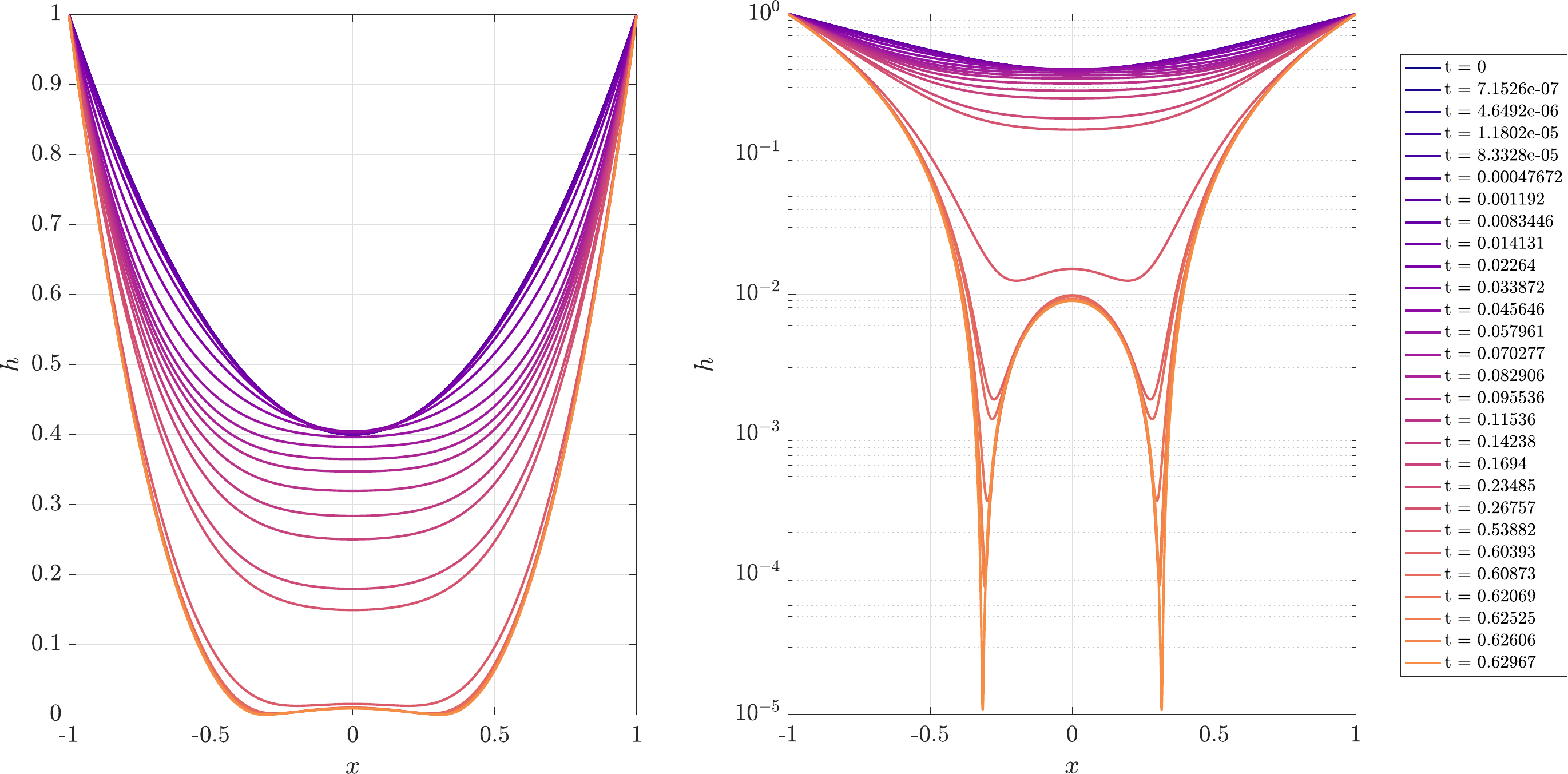}
\caption{Evolution of the filament interface for the initial condition (\ref{eq:IC_dupont}) with imposed pressure (\ref{eq:pressure}), where $p_{0} = 5/4$, $p_{f} = 5$ and $t_s=0.05$.}
\label{fig:asymmetricpde2}
\end{figure}

The notable feature of the simulations presented in figures~\ref{fig:symmetricpde1} and \ref{fig:symmetricpde2} is that the rupture is symmetric about $x=0$.  In this subsection we make a very simple change to the parameters to generate solutions that appear to rupture asymmetrically.  In particular, we again consider (\ref{eq:nondimensional})--(\ref{eq:bcs2}) with initial condition (\ref{eq:IC_dupont}) and boundary pressure (\ref{eq:pressure}), using the same parameters as in figures~\ref{fig:symmetricpde1} and \ref{fig:symmetricpde2}, except we change the final pressure $p_f$ to be $p_f=5$ instead of $p_f=2.5$.  With this change, we present numerical results in figures~\ref{fig:asymmetricpde1} and \ref{fig:asymmetricpde2}.

For the simulation in figure~\ref{fig:asymmetricpde1}, the initial condition is already in the form of a filament with an very thin neck at $x=0$.  For a short period of time, the local minimum of the thickness remains at $x=0$.  However, at some point the shape bifurcates into two minima and proceeds to rupture at two points.  This behaviour is best observed in figure~\ref{fig:asymmetricpde1}(b), where the thickness is plotted on a log scale.  Turning to figure~\ref{fig:asymmetricpde2}, while the initial filament thickness is not particularly small, the qualitative behaviour is similar to figure~\ref{fig:asymmetricpde1} in the sense that the filament thickness bifurcates at some point in time so that there are two local minima.  This situation with two local minima continues until the filament ruptures at two points in finite time.

\begin{figure}
\centering
\includegraphics[width=0.8\textwidth]{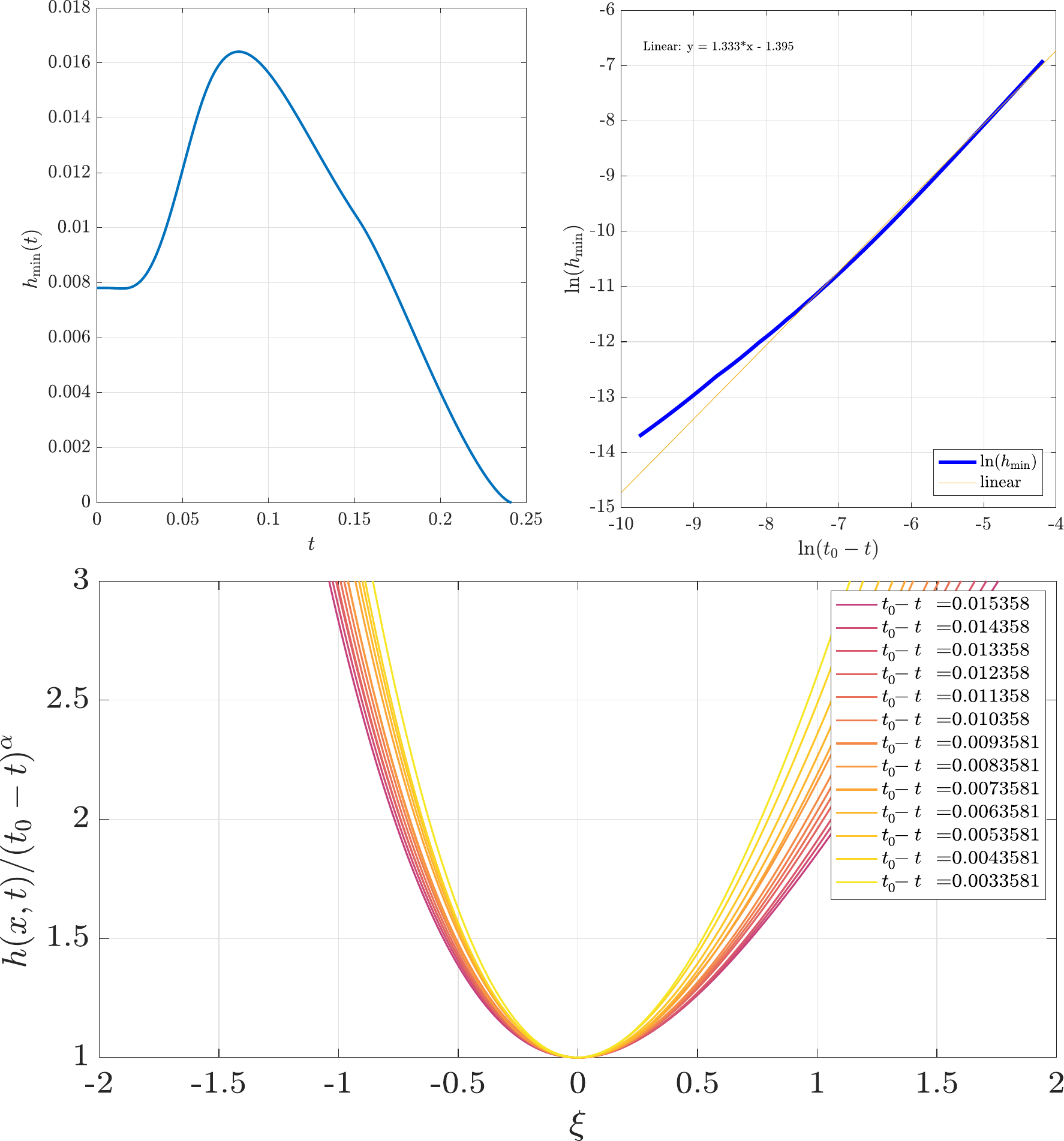}
\caption{Further results from the simulation shown in figure~\ref{fig:asymmetricpde1}.  (a) A plot of the minimum thickness $h_{\mathrm{min}}(t)$ versus time, which is used to predict the rupture time $t_0$.  The rupture location $x=x_0$ is approximated by the location of $h_{\mathrm{min}}(t)$ at $t=t_0$.  (b) The same data plotted on a log-log plot, which is used to predict the similarity parameter $\alpha$.  (c)  Given the numerically estimated $t_0$ and $\alpha$, rescaled profiles of the filament interface, $h(x,t)/(t_0-t)^\alpha$, are plotted against the similarity variable $\xi=(x-x_0)/(t-t_0)^{(1+\alpha)/4}$ using thin curves.}
\label{fig:asymmetricpde3}
\end{figure}

We wish to zoom in and observe the rates at which these rupture events are occurring.  In figure~\ref{fig:asymmetricpde3}(a), we have plotted the minimum filament thickness $h_{\mathrm{min}}(t)$ versus time for the simulation in figure~\ref{fig:asymmetricpde1}, which can be used to estimate the rupture time $t_0$.  Further, in figure~\ref{fig:asymmetricpde3}(b) we can use this value of $t_0$ to plot $\ln(h_{\mathrm{min}})$ versus $\ln(t_0-t)$.  In this example, the slope is estimated to be roughly $1.32$.  We return to part (c) of this figure below.

\section{Similarity solutions of (\ref{eq:nondimensional})}\label{sec:similarity}


We look for similarity solutions of (\ref{eq:nondimensional}) by posing the ansatz
\begin{equation}
h=(t_0-t)^\alpha f(\xi),
\quad
\xi=\frac{x-x_0}{(t_0-t)^\beta}.
\label{eq:similarityansatz}
\end{equation}
By substituting (\ref{eq:similarityansatz}) into (\ref{eq:nondimensional}) directly, we
find that
\begin{equation}
\beta=\sfrac{1}{4}(1+\alpha)
\label{eq:beta}
\end{equation}
and the similarity function $f$ must satisfy
\begin{equation}
-\alpha f+\sfrac{1}{4}(1+\alpha)\xi f'+(ff''')'=0,
\quad -\infty<\xi<\infty.
\label{eq:odef}
\end{equation}
For boundary conditions, we must ensure that $h_t$ remains finite in the far-field, thus $-\alpha f+\beta\xi f'\sim 0$, or,
\begin{equation}
f\sim C_{\pm}|\xi|^{4\alpha/(1+\alpha)}
\quad\mbox{as}\quad \xi\rightarrow \pm\infty.
\label{eq:farfieldf}
\end{equation}
Alternatively, we can infer (\ref{eq:similarityansatz})--(\ref{eq:beta}) from (\ref{eq:stretching}), where $p=4\alpha/(\alpha+1)$.

As yet, $\alpha$ is a free parameter that acts like an eigenvalue for the problem (\ref{eq:odef})--(\ref{eq:farfieldf}).  Note these equations are invariant under the stretching change of variables $\xi=\lambda\hat{\xi}$, $f=\lambda^p\hat{f}$, where $p=4$.  Together, the existence of the free parameter $\alpha$ and this invariance confirm that the similarity solution is of the second kind, which means $\alpha$ can only be determined by solving the full boundary-value problem (\ref{eq:odef})--(\ref{eq:farfieldf}).
We observe briefly that there is an exact solution of (\ref{eq:odef})--(\ref{eq:farfieldf}), namely $f=C\xi^2$, which applies for $\alpha=1$.  In original variables, this solution is $h=C(x-x_0)^2$.


We solve (\ref{eq:odef})--(\ref{eq:farfieldf}) numerically by writing (\ref{eq:odef}) as a system of four first-order equations and applying AUTO-07p
\cite{doedel2007auto}, which is a numerical continuation package designed for odes.  In fact, we generalise (\ref{eq:odef})--(\ref{eq:farfieldf}) to a family of similarity solutions of the thin film equation (\ref{eq:tfe}), depending on a parameter $m$.  Then, by continuation, we are able to begin with the exact solution $f=\xi^2$ for $\alpha=1$ which, in this more general family, is for $m=1$, and generate symmetric solutions for $m\neq 1$ along a particular branch.  This branch undergoes a fold bifurcation and returns to a solution for $m=1$ at the approximate value $\alpha=1.236$, which is the solution to (\ref{eq:odef})--(\ref{eq:farfieldf}) that we are after.  We defer further details of the scheme until section~\ref{sec:tfe}.


We have included a profile of the similarity solution with $\alpha=1.236$ in both figures~\ref{fig:symmetricode}(c) and \ref{fig:symmetricode2}(c) as a thick dashed curve.  Also included in these plots are rescaled version of the numerical solutions shown in figures~\ref{fig:symmetricpde1} and \ref{fig:symmetricpde2}, in order to compare with the similarity solutions.  In both figures, we see the time-dependent solution appear to evolve towards the similarity solution in the rupture limit.

This symmetric similarity solution with $\alpha=1.236$ has been computed very recently by King \& Bowen~\cite{king2025}, who compute various similarity solutions of the thin film equation (\ref{eq:tfe}), and Neville et al.~\cite{neville2025}, who study the rupture of a strip of active chiral fluid.
As far as we are aware, despite the very many studies of (\ref{eq:nondimensional}), there has been no previous illustration of time-dependent solutions of (\ref{eq:nondimensional}) evolving to the symmetric similarity solution in the rupture limit.


\section{The thin film equation}\label{sec:tfe}

We shall now briefly turn out attention to the thin film equation (\ref{eq:tfe}), which generalises our filament model (\ref{eq:nondimensional}) by including a parameter $m$.  The literature on (\ref{eq:tfe}) is too broad to describe here; however, it is worth noting some early studies of this model in Refs~\cite{bertozzi1996,Bertozzi1994,myers1998,Smyth1988}.  Like many other authors, Myers~\cite{myers1998} pays particular attention to $m=3$, as that is the important case for a surface-tension-driven thin film on a flat substrate (hence the name, thin film equation).

We shall not explore time-dependent solutions of (\ref{eq:tfe}), but instead focus our brief discussion on similarity solutions that govern the near-rupture limit.  In particular, we consider solutions of the form
\begin{equation}
h=(t_0-t)^\alpha f(\xi),
\quad
\xi=\frac{x-x_0}{(t_0-t)^{(1+m\alpha)/4}},
\label{eq:similarityansatz_m}
\end{equation}
where $f$ is a solution of
\begin{equation}
-\alpha f+\sfrac{1}{4}(1+m\alpha)\xi f'+(f^mf''')'=0,
\quad -\infty<\xi<\infty,
\label{eq:odef_m}
\end{equation}
\begin{equation}
f\sim C_{\pm}|\xi|^{4\alpha/(1+m\alpha)}
\quad\mbox{as}\quad \xi\rightarrow \pm\infty.
\label{eq:farfieldf_m}
\end{equation}
For the case $m=1$, (\ref{eq:similarityansatz_m})--(\ref{eq:farfieldf_m}) reduce to (\ref{eq:similarityansatz})--(\ref{eq:farfieldf}), as expected.

To solve (\ref{eq:similarityansatz_m})--(\ref{eq:farfieldf_m}) numerically, we begin by rescaling the dependent variable via
\[
f(\xi) = \left(\frac{\alpha m + 1}{4}\right)^{1/m} \hat f(\xi),
\]
so that
\[
-\gamma \hat f + \xi \hat f' + (\hat f^m\hat f''')' = 0, \qquad \gamma = \frac{4\alpha}{\alpha m+1}.
\]
This is written as the fourth-order system:
\begin{equation}
\begin{bmatrix}
\hat f \\ \hat f' \\ \hat f'' \\  q
\end{bmatrix}'
=
\begin{bmatrix}
\hat f' \\ \hat f'' \\ \D \hat f^{-m} q \\ \gamma \hat f - \xi \hat f'
\end{bmatrix}
\label{eq:selfSimODEs}
\end{equation}
where we have defined the (scaled) flux $q = \hat f^m \hat f'''$.

Assuming symmetry in $\xi$, we pose the system \eqref{eq:selfSimODEs} on a large domain $\xi \in [0,L]$, and require boundary conditions at $\xi=0$ and $\xi = L$.  At $\xi = 0$, we introduce an artificial parameter $\delta$, representing a nonzero slope at zero $\hat f'(0)$, which will allow us to move to different solution branches.  Thus we apply the boundary conditions
\[
\hat f'(0) = \delta, \qquad q(0) = 0,
\]
understanding that $\delta=0$ represents a true symmetric similarity solution.

In the far field, respresented by $\xi = L \gg 1$, we enforce the boundary condition (\ref{eq:farfieldf_m}).  A WKB analysis near this far-field condition demonstrates that it effectively imposes two constraints on the system, as two oscillatory growing modes are required to be absent \cite[e.g.]{zhang1999similarity}.  Furthermore, in our computation of similarity solutions, we eliminate the degree of freedom by taking the prefactor $C_+$ to be unity.  We thus apply the three boundary conditions:
\[
\hat f'(\xi) = \gamma \xi^{\gamma-1}, \quad
 \hat f''(\xi) = \gamma(\gamma-1)\xi^{\gamma-2}, \quad
 q = \gamma(\gamma-1)(\gamma-2)\xi^{(m+1)\gamma - 3},
\quad
\xi = L.
\]
Note that we expect five boundary conditions in total for the fourth order system as we only expect solutions for discrete values of $\gamma$.

To compute solutions we use a homotopy continuation technique.  When $\gamma = 2$ the exact solution $\hat f(\xi) = \xi^2$ satisfies the system and boundary conditions (with $\delta=0$) for any $m$.  We start with this exact solution, and $m=1$.  Then, while letting $\delta$ vary, we increase $\gamma$ until $\delta$ returns to zero.  This solution then corresponds to a symmetric similarity solution for $m=1$.

\begin{figure}
\centering
\includegraphics[width=.7\textwidth]{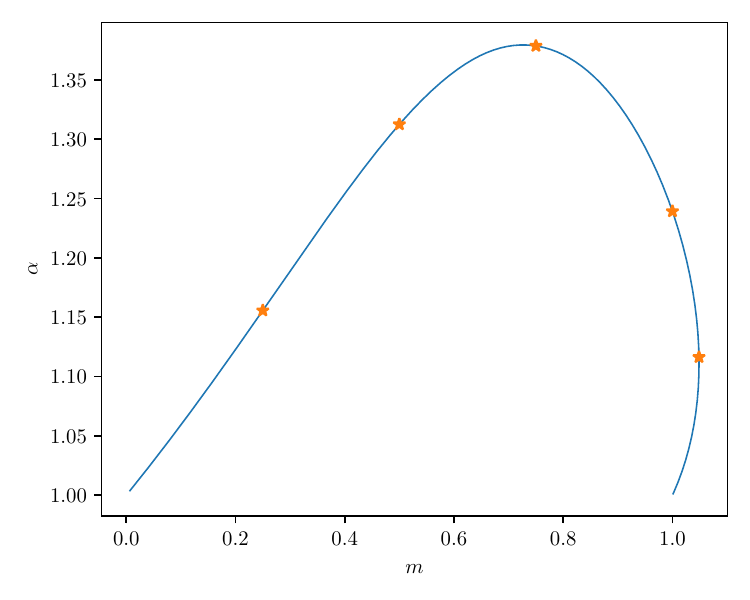}
\caption{The branch of similarity solutions of the thin film equation (\ref{eq:tfe}) calculated using the numerical continuation method described in section~\ref{sec:tfe}.}
\label{fig:branch}
\end{figure}

\begin{figure}
\centering
\includegraphics[width=.7\textwidth]{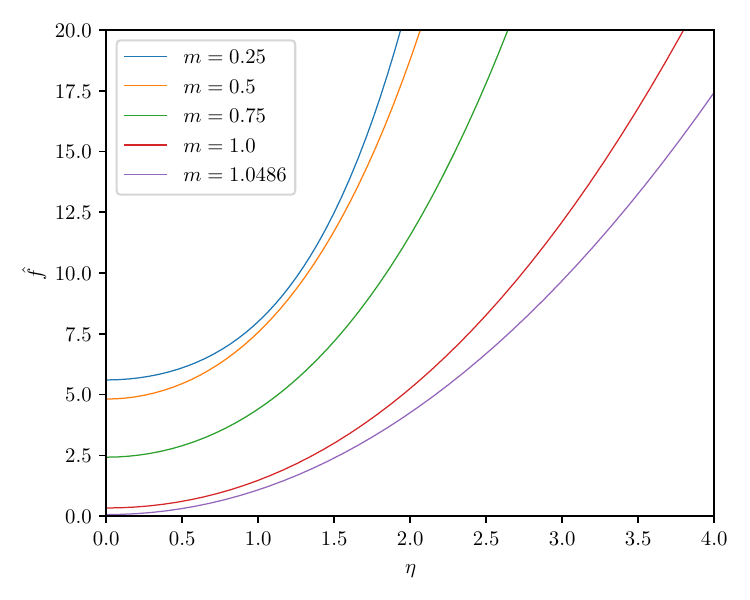}
\caption{Similarity solutions of the thin film equation (\ref{eq:tfe}) calculated at specific points on the branch depicted in figure~\ref{fig:branch}.  While illustrated here for $0<\xi<4$, the actual far-field conditions are imposed at the point $L=50$.}
\label{fig:simSols}
\end{figure}

We also trace out a branch of symmetric similarity solutions by starting with this solution and varying $m$.  The branch of solutions is depicted in figure~\ref{fig:branch}, where we have computed the value of the similarity exponent $\alpha$ from $\gamma$.  This branch of solutions is the main branch found in King \& Bowen~\cite{king2025}.  Solutions for specific values of $m$, indicated as marks on the solution branch, are depicted in figure~\ref{fig:simSols}.

Some notable features of the solution branch are as follows.  The solution branch has a turning point at $m \approx 1.049$, then loops back to the point $(m,\alpha) = (1,1)$; in this limit, similarity solutions approach the exact solution $\hat f = \xi^2$.  In the other direction, the solution branch approaches $(m,\alpha) = (0,1)$.  In this limit the governing equation becomes linear, and solutions approach the exact solution:
\[
\hat f(\xi) = 6 + \xi^4.
\]
The results in figure~\ref{fig:branch} have also been presented very recently by King \& Bowen \cite{king2025}.

Finally, we note that accurate computation of these solutions appears to be sensitive and requires a significant domain size.  The results depicted here are for $L=50$, at which point the solutions have converged for this solution branch.  It is very likely that further branches of solutions exist, and while we have found numerical evidence of these, we do not have confidence that our numerical scheme has sufficiently converged, so we do not present these results here.  A further discussion of certain other branches of solutions is provided in~\cite{king2025}.

\section{Discussion}\label{sec:discussion}

At the outset of this discussion, we emphasise that the literature on (\ref{eq:nondimensional}) and its generalisation (\ref{eq:tfe}) is incredibly vast and we have not attempted to summarise it more broadly.  Instead, we are focusing on the specific issue of filament rupture and, in particular, the form of the solution during the short time period leading up to rupture.  To create some scenarios that involve rupture, we have applied the same boundary and initial conditions as Dupont et al.~\cite{dupont1993} and computed solutions of (\ref{eq:nondimensional}) numerically.  This set-up involves a thin filament of viscous fluid in a Hele-Shaw cell subject to a pressure condition at the boundaries of the domain.  The pressure has the effect of removing fluid from the filament at a sufficiently high flow rate that the filament eventually ruptures at a finite time.

One of these examples is shown in figure~\ref{fig:asymmetricpde1}, which is like Dupont et al.'s figure 3.  Here the rupture appears to be asymmetric.  Another similar example is shown in figure~\ref{fig:asymmetricpde2}.  In contrast, the two examples in figures~\ref{fig:symmetricpde1} and \ref{fig:symmetricpde2} are different in that they appear to give rise to a symmetric rupture profile.  In all cases, we present evidence that supports the claim that the near-rupture behaviour for these time-dependent numerical solutions is described by a self-similar solution of the second kind (\ref{eq:similarityansatz})--(\ref{eq:beta}).

By solving the boundary-value problem (\ref{eq:odef})--(\ref{eq:farfieldf}) numerically, we are able to compute a symmetric similarity solution with similarity exponent $\alpha = 1.236$.  This solution has also be computed very recently by King \& Bowen~\cite{king2025} and Neville et al.~\cite{neville2025}.  By rescaling our time-dependent solutions using the similarity variables, it appears that the symmetric similarity solution with $\alpha = 1.236$ represents the appropriate symmetric near-rupture behaviour depicted in figures~\ref{fig:symmetricpde1} and \ref{fig:symmetricpde2}.

Furthermore, by generalising these solutions to hold for the thin film equation (\ref{eq:tfe}), we have been able to compute a branch of symmetric similarity solutions that depends on the parameter $m$, with $m=1$ relevant for our Hele-Shaw filament (see figure~\ref{fig:branch}).  This branch is also presented in King \& Bowen~\cite{king2025}.  However, we have not presented any asymmetric similarity solutions for the thin film equation as we are not confident that our algorithm has converged to a true solution.  Therefore, we have not been able to include a similarity solution in figure~\ref{fig:asymmetricpde3}.

We close by mentioning some open problems.  First, an obvious challenge is to compute additional families of symmetric similarity solution to extend the results in figure~\ref{fig:branch}.  Some of these are presented in King \& Bowen~\cite{king2025}, but we believe there are more branches to be determined.  Further, there may well be branches of asymmetric similarity solutions whose existence or otherwise is still a mystery.  We emphasise that this report summarises work undertaken during a three-week workshop; as such, our scope for conducting an exhaustive study of these issues was limited.

Next, for the thin filament problem we consider here, rupture is caused by a type of forcing, whereby we extract fluid from the viscous filament at the two ends of the domain.  Without this forcing, the filament does not rupture (see figure~\ref{fig:decay}, for example).  In a similar vein, Neville et al.~\cite{neville2025} has been able to generate time-dependent numerical solutions of a model that leads to symmetric similarity solutions of (\ref{eq:nondimensional}) but, again, a type a forcing is required (in that case it is to allow the fluid to be active).  It could be that there are other interesting models that force time-dependent solutions to rupture in a similar way, for example the fourth-order model
    $$
    h_t+(hh_{xxx}+h^{1/2}h_x)_x=0,
    $$
where the backward diffusion term could drive the rupture, even with periodic boundary conditions (that is, even if we apply (\ref{eq:bcs1})--(\ref{eq:bcs2}) with $P(t)\equiv 0$).  We shall consider this issue elsewhere.

Another area for future research involves studying a model that allows for the centreline of a viscous filament in a Hele-Shaw cell to be curved (remembering that our model (\ref{eq:nondimensional}) requires that the filament is symmetric about a straight centreline).  Such a model was derived recently by Dallaston and coworkers \cite{BenShachar2025,Dallaston2024}
    for the case in which the two filament interfaces are subject to a pressure difference $\Delta P$ that drives flow in the perpendicular direction to the filament centreline.  Thus, in order to write down a model that generalises (\ref{eq:pdedimensional}) to allow for a curved evolving centreline, we should set $\Delta P=0$ in Ref.~\cite{BenShachar2025,Dallaston2024}.  It would be interesting to study rupture and stability of such a model.

\vspace{3ex}
\noindent {\bf Acknowledgements:} The authors would like to thank John Lister for many fruitful discussions during the MATRIX Workshop \textit{Slow Viscous Flows with Interfaces}.

\bigskip
\bibliographystyle{spmpsci}
\bibliography{references}

\end{document}